\documentclass[twocolumn,showpacs,preprintnumbers,amsmath,amssymb]{revtex4}

\usepackage{graphicx}
\usepackage{dcolumn}
\usepackage{bm}

\begin{document}

\title{Quantum Vacuum and the Structure of  \emph{Empty}  Space-Time}

\author{Ashot Gevorkyan}
 \altaffiliation{Institute for Informatics and Automation Problems/Institute of Chemical Physics,
  NAS of RA}
 \email{g_ashot@sci.am}

\date{\today}

\begin{abstract}
We have considered the possibility of formation a massless particles with spin 1 in
the region of negative energies, within the framework of the Weyl type equation for neutrinos.
It is proved that, unlike quantum electrodynamics, the developed approach allows
in the \emph{ground state} the formation of such stable particles. The structure and properties
of this vector-boson are studied in detail.  The problem of entangling two vector
bosons with projections of spins +1 and -1 and, accordingly, the formation of a
zero-spin boson is studied within the framework of a complex stochastic equation of
the Langevin type. The paper discusses the structure of the Bose particle of a scalar
field and the space-time's properties of an empty space (\emph{quantum vacuum}).

\end{abstract}

\maketitle
\section{Introduction}
\label{01}
From a mathematical and philosophical point of view, the vacuum can be comparable with
the region of absolutely empty space or, which is the same, with the region of the
space where there are no massive particles and fields. From the physical point of view,
the real vacuum is empty only "\emph{on average}". Moreover, as it is known, due to
the principle of quantum-mechanical uncertainty, there is not any way screening a certain
area of space to exclude the appearance of virtual particles and fields in it.
The Lamb Shift \cite{Lamb}, Casimir effect \cite{Casimir}, Unruh effect \cite{Unru},
anomalous magnetic moment of electron \cite{Proh}, Van der Waals forces \cite{Van},
Delbr\"{u}ck scattering \cite{Del}, Hawking radiation \cite{Haw}, the cosmological constant
problem \cite{Weinb,Wein}, vacuum polarization at weak electromagnetic fields \cite{Ash1,Ash2}
- here is an incomplete list of phenomena, part of which has been experimentally discovered.
 All of them are conditioned by the
physical vacuum or, more accurately, by a \emph{quantum vacuum} (QV). The issues of
quintessence (dark energy) and cosmic acceleration often are discussed in the framework of QV
theories, which necessarily include  scalar fields \cite{Wein}. The properties of
a quantum vacuum can be studied within the framework of quantum field theory (QFT),
ie quantum electrodynamics and quantum chromodynamics (QCD). Note, that QFT could accurately
describe QV if it were possible to exactly summarize the infinite series of perturbation
theories, that is typical of field theories. However, it is well-known that  the perturbation
theory for QFT breaks down at low energies (for example, QCD or the
BCS theory of superconductivity) field operators may have non-vanishing vacuum expectation
values called condensates. Moreover, in the Standard Model precisely the non-zero vacuum
expectation value of the Higgs field, arising from spontaneous symmetry breaking, is
the principled mechanism allowing to acquire masses of other fields of theory.

To overcome these difficulties and to conduct a consistent and comprehensive study of
the quantum vacuum, we developed a nonperturbative approach based on a system of complex
stochastic equations of the Langevin type describing the motion of a massless particle
with spin 1 in a random environment of similar particles.

The main objectives pursued by this study are as follows:

a. to study the possibility of formation a stable bound state of a massless particle
with spin 1 (vector boson) in the region of negative energies,

b. to investigate the possibility of forming a massless particle with zero spin as a
result of entangling the ground states of two vector bosons, respectively, with the
projections of the spins 1 and -1,

c. to justify of the possibility the formation of a scalar field as a result of
relaxation and Bose-Einstein (BEC) condensation of vector fields. To study the features
the wave state of the boson of a scalar field, in particular, the possibility of its
decay and spontaneous transitions to other vacuum and extravacuum states.

\section{Quantum motion of a photon in empty space}
The questions of correspondence between the Maxwell equation and the equation of quantum mechanics
was of interest to many researchers at the dawn of the development of quantum theory \cite{Opp,Mol,Wein1}.

As shown \cite{BIALYNICKI}, the quantum motion of a photon in a vacuum can be considered
within the framework of a wave function representation, writing it analogously to the Weyl
equation for a neutrino in a vector form:
\begin{eqnarray}
\partial_{t}\bm{\Psi^{+}}(\textbf{r},t)-c_0\bigl(\textbf{S}\cdot \bm\nabla\bigr) \bm{\Psi^{+}}(\textbf{r},t)=0,
\qquad\qquad\qquad\,\,\,\nonumber\\
\partial_{t}\bm{\Psi^{-}}(\textbf{r},t)+c_0\bigl(\textbf{S}\cdot \bm\nabla\bigr) \bm{\Psi^{-}}(\textbf{r},t)=0,
\qquad \partial_{t}\equiv\partial/\partial t,
\label{1.01a}
\end{eqnarray}
where $c_0$  denotes the speed of light in an empty $ 4D $ Minkowski space-time,
$\bm{\Psi^{+}}(\textbf{r},t)$ and $\bm{\Psi^{-}}(\textbf{r},t)$ denote the photons wave functions
of both helicities, corresponding to left-handed and right-handed polarizations. In addition,
in (\ref{1.01a}) the set of matrices $\textbf{S}=(S_x,S_y,S_x)$ describes infinitesimal
rotations of particles with spin $\pm1$, respectively:
\begin{eqnarray}
S_x=\left[
  \begin{array}{ccc}
    0 & 0 & 0\\
    0 & 0 & -i \\
    0 & i & 0 \\
  \end{array}
\right],
\qquad
S_y=\left[
  \begin{array}{ccc}
    0 & 0 & i\\
    0 & 0 & 0 \\
    -i & 0 & 0 \\
  \end{array}
\right],\,\,\,
\nonumber\\
S_z=\left[
  \begin{array}{ccc}
    0 & -i & 0\\
    i & 0 & 0 \\
    0 & 0 & 0 \\
  \end{array}
\right].\qquad\qquad\quad\,\,\,\,\,
\label{1.01cy}
\end{eqnarray}
 Recall, that in (\ref{1.01a}) the absence of electrical and magnetic charges provides the following conditions:
 \begin{equation}
 \nabla \cdot \bm{\Psi^{\pm}}(\textbf{r},t)=0.
\label{1.01}
\end{equation}
If we represent the wave function in the form:
\begin{eqnarray}
 \bm{\Psi^{\pm}}(\textbf{r},t)=\frac{1}{\sqrt{2}}\biggl\{\frac{\textbf{D}(\textbf{r},t)}{\sqrt{\epsilon_0}}
 \pm i\frac{\textbf{B}(\textbf{r},t)}{\sqrt{\mu_0}}\biggr\},\,\, c_0=\frac{1}{\sqrt{\mu_0\epsilon_0}},
\label{1.01b}
\end{eqnarray}
then from the equations (\ref{1.01a}) and (\ref{1.01}) it is easy to find Maxwell's equations in
an ordinary vacuum or in empty space:
\begin{eqnarray}
\partial_t \textbf{D} -\nabla\times\textbf{H} =0,\qquad
 \nabla\cdot\textbf{E}=0,\nonumber\\
\partial_t \textbf{B} +\,\nabla\times\textbf{E}=0,\qquad
 \nabla\cdot\textbf{H}=0,
 \label{1.01c}
\end{eqnarray}
where $\epsilon_0$ and $\mu_0$ describe the dielectric and magnetic constants of the vacuum, respectively.
It is important to note that the dielectric and magnetic constants provide the following equations:
$$\textbf{D}=\epsilon_0\textbf{E}, \qquad \textbf{B}=\mu_0\textbf{H}.$$
Recall that the only difference between the equations (\ref{1.01a}) and (\ref{1.01c}) is
that the Maxwell equations system does not take into account the spin of the photon, which
will be important for further constructions. Since the refractive indices $ \epsilon_0 $ and
$\mu_0 $ are constants that do not depend on external fields, and characterize the state of
\emph {unperturbed} or ordinary vacuum, a reasonable idea arises, namely, to consider a
vacuum or, more accurately, QV, as some energy environment with \emph{unusual properties and structure}.

\subsection{Vector fields and their fundamental particle}
 Let us make the following substitutions in the equations (\ref{1.01a}):
\begin{equation}
c_0\mapsto c(x,y,z,\tau), \qquad t\mapsto \tau(t),
\label{2.0a1}
\end{equation}
where $c$ is the velocity of propagation of the field in the structural particle,
which differs from the speed of light $c_0$ in a vacuum, $\tau$  is the
chronological parameter of a closed quantum system (\emph{internal time}), which,
naturally, must be quasi-periodic, and in the most general case can be represented as:
\begin{equation}
 \tau(t)=T(t)\sin(t/T(t)),\qquad -\infty<t<+\infty,
\label{2.0b1}
\end{equation}
where $ T(t)>0$ is a bounded function-quasiperiod. \\

\textbf{Proposition}. \emph{The equations system (\ref{1.01a})-(\ref{1.01}),
taking into account the  remarks (\ref{2.0a1}) and  (\ref{2.0b1}) describes
the quantum states  of the massless  particles with spin projections $\pm1$.}

The particle of the vector field  in depending on the value of the spin projection,
can be described by one of the following sets of wave functions:
 \begin{eqnarray}
\label{1.01cz}
\bm\psi^{+}(\textbf{r},\tau)=\left[
  \begin{array}{ccc}
    \psi^{+}_x(\textbf{r},\tau)\\
   \psi^{+}_y(\textbf{r},\tau)\\
   \psi^{+}_z(\textbf{r},\tau) \\
  \end{array}
\right],\,\,
\bm\psi^{-}(\textbf{r},\tau)=\left[
  \begin{array}{ccc}
    \psi^{-}_x(\textbf{r},\tau)\\
   \psi^{-}_y(\textbf{r},\tau)\\
   \psi^{-}_z(\textbf{r},\tau) \\
  \end{array}
\right],
\end{eqnarray}
which satisfy equations of the type (\ref{1.01a})-(\ref{1.01}) taking into account the above remarks.

Obviously, in the \emph{bound state}, the 4D-interval of the propagated signal will be
zero, and the points of the Minkowski space (events) are related to the relation, like
the light cone:
\begin{equation}
s^2 =c^2\tau^2-r^2=0, \qquad r^2=x^2+y^2+z^2.
 \label{1.02k}
\end{equation}

Thus, the key question arises: can such equations describe the spatial localizations of massless
fields, characteristic of a stable structured formations called a particles?

Substituting (\ref{1.01cz}) into (\ref{1.01a}) and taking into account (\ref{1.01cy}), we can find the following
two independent system of first order partial differential equations:
\begin{eqnarray}
ic^{-1}\partial_\tau \psi^{\pm}_x=\pm\partial_y \psi^{\pm}_z\mp\partial_z \psi^{\pm}_y,
\nonumber\\
ic^{-1}\partial_\tau \psi^{\pm}_y=\pm\partial_z\psi^{\pm}_x\mp\partial_x \psi^{\pm}_z,
\nonumber\\
ic^{-1}\partial_\tau \psi^{\pm}_z=\pm\partial_x\psi^{\pm}_y\mp\partial_y \psi^{\pm}_x.
\label{1.02tab}
\end{eqnarray}
In addition, the following relation can be found between the fields, taking into account the
equation (\ref{1.01}):
\begin{eqnarray}
  \partial_x{\psi}^{\pm}_x+  \partial_y{\psi}^{\pm}_y+ \partial_z{\psi}^{\pm}_z=0.
\label{1.02ta}
\end{eqnarray}

Since particles with spins $+1$ and $-1$ are symmetric in any sense, it is sufficient to
study in detail only the wave function describing a particle with spin $+1$.

Using the equations (\ref{1.02tab}) and (\ref{1.02ta}) we can obtain the  systems of second
order partial differential equations for vacuum fields:
\begin{widetext}
\begin{eqnarray}
\square \psi^{+}_x={c}^{-1}{c_{,y}}\bigl(\partial_x \psi^{+}_y-\partial_y \psi^{+}_x\bigr)-
{c}^{-1} {c_{,z}} \bigl(\partial_z \psi^{+}_x-\partial_x \psi^{+}_z \bigr)- c_{,\tau}c^{-3}\partial_\tau\psi^{+}_x,
 \nonumber\\
\square\psi^{+}_y={c}^{-1}{c_{,z}}\bigl(\partial_y \psi^{+}_z-\partial_z \psi^{+}_y\bigr)-
{c}^{-1}{c_{,x}}\bigl(\partial_x \psi^{+}_y-\partial_y \psi^{+}_x\bigr)-c_{,\tau}c^{-3}\partial_\tau\psi^{+}_y,
\nonumber\\
\square\psi^{+}_z= {c}^{-1}{c_{,x}}\bigl(\partial_z \psi^{+}_x-\partial_x \psi^{+}_z\bigr)-
{c}^{-1}{c_{,y}}\bigl(\partial_y \psi^{+}_z-\partial_z \psi^{+}_y\bigr)-c_{,\tau}c^{-3}\partial_\tau\psi^{+}_z,
\label{1.02t}
\end{eqnarray}
\end{widetext}
where $\square=\triangle-c^{-2}\partial_\tau^2$ denotes the D'Alembert operator, $\bigtriangleup$ is
the Laplace operator,  $c_{,\zeta}=\partial c/\partial \sigma$ and $\zeta=(x,y,z,\tau)$.
To determine the explicit form of equations (\ref{1.02t}), we need to calculate the derivatives $c_{,\sigma}$.
Using the equation (\ref{1.02k}), we can find:
\begin{equation}
c_{,\tau}=-\frac{c^2}{r},\quad c_{,x}=\frac{cx}{r^{2}}, \quad c_{,y}=\frac{cy}{r^{2}}, \quad
 c_{,z}=\frac{cz}{r^{2}}.\label{1.02zt}
\end{equation}
To continue the study of the problem, we need to reduce the system of equations (\ref{1.02t})
to the \emph{canonical form}, when the field components are separated, and each of them is
described by an autonomous equation.

Taking into account the circumstance that in this problem all fields are symmetric,
the following additional conditions can be imposed on the field components:
\begin{eqnarray}
 (c_{,z}- c_{,y})\partial_\tau \psi^{+}_x=c_{,z}\partial_\tau\psi_y^+-c_{,y}\partial_\tau\psi_z^+,
\nonumber\\
 (c_{,x}- c_{,z})\partial_\tau\psi^{+}_y=c_{,x}\partial_\tau\psi_z^+-c_{,z}\partial_\tau\psi_x^+,
\nonumber\\
(c_{,y}- c_{,x})\partial_\tau\psi^{+}_z=c_{,y}\partial_\tau\psi_x^+-c_{,x}\partial_\tau\psi_y^+.
 \label{1.02ak}
\end{eqnarray}
It is easy to verify that these conditions are symmetric with respect to the components of
the field and are given on the hypersurface of four-dimensional events.
Further, using the conditions (\ref{1.02ak}), the system of equations (\ref{1.02t}), can be reduced
to the following canonical form:
\begin{eqnarray}
 \bigl\{\square +[i(c_{,z}-c_{,y})+c_{,\tau}c^{-1}]c^{-2}\partial_\tau\bigr\}\psi^{+}_x=0,
\nonumber\\
\bigl\{\square+[i(c_{,x}-c_{,z})+c_{,\tau}c^{-1}]c^{-2}\partial_\tau\bigr\}\psi^{+}_y=0,
\nonumber\\
\bigl\{\square + [i(c_{,y}-c_{,x})+c_{,\tau}c^{-1}]c^{-2}\partial_\tau\bigr\}\psi^{+}_z=0.
 \label{1.02zkl}
\end{eqnarray}

Note that unperturbed QV fields must satisfy the conditions of autonomy, that imply the
separation of spatial and temporal  components of the wave function. Proceeding from this,
it is convenient to represent the wave function in the form:
 \begin{equation}
 \psi_\sigma^+(\textbf{r},\tau)=\exp{\Bigl(\frac{\mathcal{E}_\sigma \tau}{\hbar}\Bigr)}\phi_\sigma^+(\textbf{r}),
 \qquad  \sigma=x,y,z,
\label{1.02a}
\end{equation}
where $\mathcal{E}_\sigma<0$ is the energy of the one mode of QV field. It is obvious that the
symmetry of the problem implies the equality of the modes energies
$\mathcal{E}_x=\mathcal{E}_y=\mathcal{E}_z=\mathcal{E}<0$.

Finally, substituting (\ref{1.02a}) into (\ref{1.02zkl}), taking into account
(\ref{1.02zt}), we can get the following system of equations:
\begin{eqnarray}
\Bigl\{\triangle+\lambda\Bigl[-\lambda-\frac{1}{r}+i\frac{z-y}{ r^2}\Bigr]\Bigr\}\phi_x^+(\textbf{r})
=0,
\nonumber\\
\Bigl\{ \triangle + \lambda\Bigl[-\lambda-\frac{1}{r}+ i\frac{x-z}{ r^2}\Bigr]\Bigr\}\phi_y^+(\textbf{r})
=0,
\nonumber\\
\Bigl\{\triangle +\lambda\Bigl[-\lambda-\frac{1}{r}+i\frac{y-x}{r^2}\Bigr]\Bigr\}
 \phi_z^+(\textbf{r})=0,
\label{1.02b}
\end{eqnarray}
where $\lambda=(\mathcal{E}/c\hbar)<0$, in addition, parameter $|\lambda|$ has
dimensionality of inverse distance.
\vspace{7mm}
\section{The wave function of a massless particle with  spin $+1$  }

Representing the wave function in the form:
\begin{equation}
\phi_x^+(\textbf{r})=\phi_x^{+(r)}(\textbf{r})+i\phi_x^{+(i)}(\textbf{r}),
\label{3.020}
\end{equation}
from the first equation of the system (\ref{1.02b}), we can get the following two equations:
\begin{eqnarray}
\Bigl\{\triangle - \lambda\Bigl(\lambda+\frac{1}{r}\Bigr)\Bigr\}
\phi_x^{+(r)}(\textbf{r})-\lambda\frac{z-y}{r^2}\phi_x^{+(i)}(\textbf{r})
=0,
\nonumber\\
\Bigl\{\triangle - \lambda\Bigl(\lambda+\frac{1}{r}\Bigr)\Bigr\}\phi_x^{+(i)}(\textbf{r})+
\lambda\frac{z-y}{r^2}\phi_x^{+(r)}(\textbf{r})=0.
\label{3.02a}
\end{eqnarray}
It can be verified that by using simple substitutions of unknowns  $\phi_x^{+(r)}(\textbf{r})
\mapsto \phi_x^{+(i)}(\textbf{r})$ and $\phi_x^{+(i)}(\textbf{r}) \mapsto -\phi_x^{+(r)}(\textbf{r})$,
these equations pass into each other. The latter means that the solutions in modulus are
equal and differ only in sign. In other words, the symmetry properties mentioned above
make it possible to obtain two independent equations of the form:
\begin{eqnarray}
\Bigl\{ \triangle - \lambda\Bigl[\lambda+\frac{r-y+z}{ r^2}\Bigr]\Bigr\}\phi_x^{+(r)}(\textbf{r})=0,
\nonumber\\
\Bigl\{ \triangle - \lambda\Bigl[\lambda+\frac{r+y-z}{ r^2}\Bigr]\Bigr\}\phi_x^{+(i)}(\textbf{r})=0.
\label{3.02b}
\end{eqnarray}

Now we analyze the possibility of solving for the term $ \phi_x^{+(r)}(\textbf{r}) $ in the
form of a localized state. Let us consider the following equation:
\begin{equation}
\label{3.02ab}
r-y+z=\mu r,
\end{equation}
where $\mu$ is a some parameter. The  changing range of this parameter will be  defined below.

Substituting (\ref{3.02ab}) into the first equation of the system (\ref{3.02b}), we get:
\begin{equation}
\label{3.03k}
\Bigl\{\triangle + \bigl[-\lambda^2-\lambda\mu{r}^{-1}\bigr]\Bigr\}\phi_x^{+(r)}(\textbf{r})=0.
\end{equation}
To continue the research, it is useful to rewrite the equation (\ref{3.03k}) in the
spherical coordinates system $(x,y,z)\mapsto(r, \theta, \varphi) $:
\begin{widetext}
\begin{equation}
\biggl\{\frac{1}{r^{2}}\Bigl[\frac{\partial}{\partial r}\Bigl(r^2\frac{\partial}{\partial r}\Bigr)
+\frac{1}{\sin^2\theta}\frac{\partial^2}{\partial\varphi^2}+\frac{1}{ \sin\theta}
\frac{\partial}{\partial\theta}\Bigl(\sin\theta
\frac{\partial}{\partial\theta}\Bigr)\Bigr]
-\lambda^2\Bigl[1- \frac{ \mu(\theta,\varphi)}{\lambda r} \Bigr]\biggr\}\phi_x^{+(r)}=0.
\label{3.03a}
\end{equation}
\end{widetext}
Using for the wave function the representation:
\begin{equation}
\phi_x^{+(r)}(\textbf{r})=\Lambda(r)Y(\theta,\varphi),
\label{3.03b}
\end{equation}
we can conditionally separate the variables in the equation (\ref{3.03a}) and write it
in the form of the following two equations:
\begin{eqnarray}
r^2\Lambda ^{''}+2r\Lambda ^{'}+\bigl[-\lambda^2 r^2-\lambda\mu(\theta,\varphi)r-\nu\bigr]\Lambda =0,
\label{3.02bt}
\end{eqnarray}
and, respectively;
\begin{eqnarray}
\frac{1}{\sin\theta}\Bigl\{
\frac{1}{\sin\theta}\frac{\partial^2}{\partial\varphi^2}+\frac{\partial}{\partial\theta}\Bigl(\sin\theta
\frac{\partial}{\partial\theta}\Bigr)\Bigr\}Y+\nu Y=0,
\label{3.02bt'}
\end{eqnarray}
where $\Lambda^{'}=d\Lambda/dr$ and $\nu$ is a constant, which is convenient to represent in the
form $\nu=l(l+1)$ and $l=0,1,2...$ Recall that the conditional separation of variables means
to impose an additional  condition on the function $\mu(\theta,\varphi)=const$.
Writing equation (\ref{3.02ab}) in spherical coordinates, we obtain the following trigonometric equation:
\begin{equation}
\label{3.02abc}
\mu(\theta,\varphi)=1-\sin\theta\sin\varphi +\cos\theta.
\end{equation}
As the analysis of the equation (\ref{3.02abc}) shows, the range of variation of the parameter
$\mu$ for real angles is $\mu\in [(1-\sqrt{2}), (1+\sqrt{2})]$.

The solution of the equation (\ref{3.02bt'}) is well known, these are spherical Laplace
functions $Y_{l}^{m}(\theta,\varphi)$ and $m=0,\pm1,...,\pm l$.

As for the equation (\ref{3.02bt}), we will solve it for  a fixed value
$\mu$, which is equivalent to the plane cut of the three-dimensional solution.
In particular, we will seek a solution $\Lambda(r)$ tending to finite value for $r\to 0 $ and,
respectively, to zero at $r\to\infty$.

For a given  parameter $\mu_0$,  we can write the equation (\ref{3.02bt}) in the form:
\begin{eqnarray}
\frac{d^2\Lambda}{d\varrho^2}+\frac{2}{\varrho}\frac{d\Lambda}{d\varrho}+\Bigl[-\beta^2 +\frac{2}{\varrho}-
\frac{l(l+1)}{\varrho^2}\Bigr]\Lambda=0,
\label{3.02btz}
\end{eqnarray}
where $\varrho=r/a_p$ is a dimensionless distance, $a_p=2/(|\lambda|\mu_0)$
denotes the characteristic spatial dimension of a hypothetical massless Bose particle with spin 1
and $\beta=2/\mu_0$. As can be seen, this equation describes the radial wave function of a
hydrogen-like system \cite{Land}. Recall that it can have a solution describing the bound state if the
following condition is satisfied:
\begin{equation}
\label{3.03}
n_r+l+1=n+l=\frac{1}{\beta},\qquad n_r=0,1,2...,
\end{equation}
where $ n_r $ is the radial quantum number, and $n$, respectively, is the principal quantum number.\\
It is easy to see that there is only one integer $\mu_0 =2$ for which $\beta=1 $ and,
accordingly, the equation (\ref{3.03}) has a solution. In other words, the equality $\mu_0= 2$
is a  necessary \emph{quantization condition}, which, with consideration (\ref{3.02abc}), is equivalent
to the trigonometric equation:
\begin{equation}
\label{3.03b}
1+\sin\theta\sin\varphi -\cos\theta=0.
\end{equation}
In particular, from this condition it follows that the $\Lambda(r) $  solution of the equation (\ref{3.02btz})
is localized on the plane  $S^r_x(\theta,\varphi)\cong(-Y,Z)$ (see FIG. 1).
\begin{figure}
\includegraphics[width=55mm]{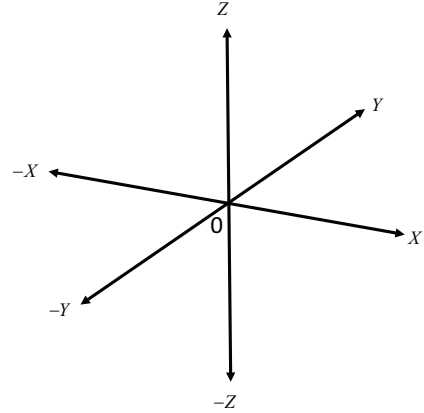}
\caption{\label{fig:epsart} The coordinate system $\{X,Y,Z\}$ divides the 3D space into eight 
spatial regions using twelve planes. The boson of a vector field with spin projection +1 is a
2D-string (brane) consisting of six components localized on the following planes $\phi^+_x[(-Y,Z)\cup(Y,-Z)],$
$\phi^+_y[(-X,Z)\cup(X,-Z)]$ and $\phi^+_z[(-X,Y)\cup(X,-Y)]$, respectively.}
\end{figure}
Proceeding from the foregoing, we can write  the solution of equation (\ref{3.02btz}) on this plane:
$$
\phi^r_x(r,\theta,\varphi)=\Lambda_{nl}(r)Y_l^m(\theta,\varphi),
$$
 which, subject to the condition (\ref{3.03b}), has a solution only for the "\emph{ground state}":
\begin{equation}
\label{3.03a}
\Lambda_{10}(r)=\frac{C}{\sqrt{a_p^3}}\,e^{-r/a_p},\qquad Y_0^0(\theta,\varphi)=\frac{1}{2}\sqrt{\frac{1}{\pi}},
\end{equation}
where $C$ denotes a constant, which will be defined below from the normalization condition
of the wave function. It is easy to see that the additional condition $\mu(\theta,\varphi)=
\mu_0=2$ does not lead to a contradiction between the equations (\ref{3.02bt}) and (\ref{3.02bt'}),
and the solution (\ref{3.03a}) is the only solution of the first equation in the system
(\ref{3.02a}) in the region of negative energies.

The imaginary part of the wave function  is calculated similar way, but in this case we
obtain following underdetermined  trigonometric equation:
\begin{equation}
\label{3.03ba}
1-\sin\theta\sin\varphi+\cos\theta=0,
\end{equation}
which determines the plane  $S^i_x(\theta,\varphi)\cong(Y,-Z)$ on which the solution of the
\emph{ground state} is localized.

Similar solutions are obtained for the projections of the wave function $\phi^+_y$ and
$\phi^+_z$. In this case, however, the components of the wave function $\bm\phi^+(\textbf{r})$
are localized on the planes  $\{S^{r}_y (\theta, \varphi)\cong(-X,Z),\,\,
S^{i}_y(\theta,\varphi) \cong(X,-Z)\}$ and $\{S^{r}_z (\theta,\varphi)\cong(X,-Y),\,
\,S^{i}_z(\theta, \varphi)\cong(-X,Y)\}$, accordingly. So, the  orient of first pair of planes
$(S^{r}_y ,\,\,S^{i}_y)$ are determined by the equations:
\begin{equation}
\label{3.04}
1\pm\cos\theta\sin\varphi\mp\cos\theta=0,
\end{equation}
while, the orientations of the second pair of planes $(S^{r}_z ,\,\,S^{i}_z)$ are given by the equations:
\begin{equation}
\label{3.04a}
1\pm\sin\theta\sin\varphi\mp\cos\theta\sin\varphi=0.
\end{equation}
It should be noted that the coefficient of the radial wave function (see (\ref{3.03a})) is chosen so that the total
vector wave function $\bm\psi^+$ of the "\emph{ground state}" is normalized to unity.

Now we can write down the normalization condition for the total wave function for the vector boson:
\begin{equation}
\label{3.04b}
J=\int \bm\phi^+\bigl(\bar{\bm\phi}^+\bigr)^TdV=1, \qquad dV=dxdydz,
\end{equation}
where $\bigl(\bar{\bm\phi}^+\bigr)^T=\bigl(\bar{\bm\phi}^+_x,\bar{\bm\phi}^+_y,\bar{\bm\phi}^+_z\bigr)$
is the transposed vector.\\
We can represent the integral (\ref{3.04b}) as a sum of three terms:
\begin{equation}
\label{3.04ab}
J=\int\phi^+_x\bar{\phi}^+_xdV+\int\phi^+_y\bar{\phi}^+_ydV+\int \phi^+_z\bar{\phi}^+_zdV.
\end{equation}
It is obvious that all terms in the expression (\ref{3.04ab}) are equal and,
therefore, each of them is equal to 1/3.

As an example, consider the first summand. Taking into account that the wave function
can be represented in the form $\phi^{+}_x=\phi^{+ (r) }_x + i\phi^{+ (i)}_x $,  and
its components  $\phi^{+ (r) }_x $ and $\phi^{+ (i)}_x $, are localized in a region
that is the union of two  quaternary planes ${(-Y,Z)\cup(Y,-Z)}\subset(Y, Z)$,  we can write:
\begin{widetext}
\begin{equation}
\label{3.05}
 \int\phi^+_x\bar{\phi}^+_xdV=\int\bigl(|\phi^{+(r)}_x|^2+|\phi^{+(i)}_x|^2\bigr)dV
 =2\int|\phi^{+(r)}_x|^2dV=2\int|\phi^{+(i)}_x|^2dV=\frac{a_p}{2\pi}\int_{(-Y, Z)}|\Lambda_{10}(\rho)|^2 dydz=1/3.
\end{equation}
\end{widetext}
where $\varrho=r(0,y,z)= \sqrt{y^2+z^2}$ denotes radius-vector $r$ on the plane $(-Y, Z)$,
in addition, in calculating the integral, we assume that the wave function in perpendicular to the plane
$(-Y, Z)$  direction $x$ is the Dirac delta function.

Taking into account (\ref{3.03a}), we can accurately calculate the integral in the expression  (\ref{3.05}):
\begin{equation}
\label{3.05a}
\int_{(-Y, Z)}|\Lambda_{10}(\rho)|^2dydz=\frac{\pi}{8a_p}C^2.
\end{equation}
Now using (\ref{3.05})  and (\ref{3.05a}), we can determine the normalization constant of the wave function
 (\ref{3.03a}), which is equal to $C= 4/\sqrt{3} $.   Note that similarly we can find a solution of the boson wave function
 with the spin projection -1  (see Appendix A).
\begin{figure}
\includegraphics[width=70mm]{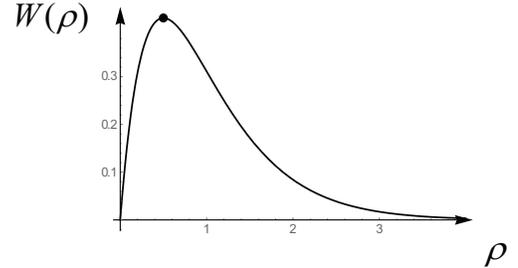}
\caption{The distance $\rho_0=1/2$, or more precisely $ \varrho_0=a_p /2$, at which the maximum
value of the amplitude of the \emph{brane} probability is reached.}
\end{figure}

Finally, we turn to the characteristic scale of the \emph{brane}, which, as follows from the
quantization condition (see (\ref{3.02btz})-(\ref{3.03})), is defined by the equation:
\begin{equation}
a_p=\frac{1}{|\lambda|}=\frac{c\hbar }{|\mathcal{E}|},
\label{3.05at}
\end{equation}
where $c$ is the \emph{speed of propagation of a wave in the brane}, which can significantly
differ from the speed of light in a vacuum. Unfortunately, it is impossible to determine
$a_p$ from equation (\ref{3.02btz}). Nevertheless, if we assume that the spatial scale \emph{brane} has
the Planck length $l_P$, ie, $a_p=l_P= 1,616,229 (38) \cdot 10^{-33}sm $, and the  energy particle is $-1MeV$,
then the propagation velocity of the wave in the \emph{brane} will be equal to $c=3,86\cdot  10^{-13}sm/s $.

Using the solution of the \emph{ground state}, on the corresponding planes, we can calculate the
location of the maximum probability of one of the projections of the \emph{brane}. The probability
of finding the maximum amplitude of the projection of a \emph{brane} wave into an element of the
area $ dS=rdrd\vartheta$ can be determined as follows:
\begin{equation}
dW(\rho)={C'}^2e^{-2\rho}{\rho}d{\rho}d\vartheta,\qquad C'=C/\sqrt{a_p^3},
\label{3.06t}
\end{equation}
where $\vartheta\in[0,\pi/2].$\\
Investigating the expression $dW/d\rho $, it is easy to find that for the value
$\rho=1/2$ and, respectively, for $r(0,x,y) =\varrho= a_p/2$, the probability distribution
has a maximum  FIG 2.
\emph{Proposition is proved.}

\section{Formation of a scalar field }
Let us consider QV, taking into account the fluctuations of the vector fields. As the
basic equations  we will use   the system of stochastic matrix equations of the Langevin type:
\begin{eqnarray}
\partial_{t}\bm{\breve{{\psi}}^{+}}(\textbf{r}_+,t)-c\bigl(\textbf{S}\cdot \bm\nabla\bigr) \bm{\breve{{\psi}}^{+}}(\textbf{r}_+,t)=\bm\eta^+(s_+),
\nonumber\\
\partial_{t}\bm{\breve{{\psi}}^{-}}(\textbf{r}_-,t)+\,c\bigl(\textbf{S}\cdot \bm\nabla\bigr) \bm{\breve{{\psi}}^{-}}(\textbf{r}_-,t)=\bm\eta^-(s_-),
\label{4.02t}
\end{eqnarray}
and also the equations:
\begin{eqnarray}
\nabla\bm{\breve{{\psi}}^{+}}(\textbf{r}_+,t)=0,
\qquad
\nabla\bm{\breve{{\psi}}^{-}}(\textbf{r}_-,t)=0,
\label{4.02k}
\end{eqnarray}
where $\textbf{r}_\pm$ denotes the radius-vector of corresponding vector particles, $t\in(-\infty,+\infty)$
is a usual time. In equation (\ref{4.02t}) complex stochastic generators $\bm\eta^\pm(s_\pm)=
(\eta_x^\pm,\eta_y^\pm,\eta_z^\pm)$, describing random charges and currents arising in 4D-intervals  $ds^2_\pm=c^2dt^2-dx^2_\pm-dy^2_\pm-dz^2_\pm$.

For further research, it is useful to write these equations in matrix form:
$$
\left[
\begin{array}{ccc}
\vspace{2mm}
ict & -z_+ & y_+\\
\vspace{2mm}
z_+ & ict & -x_+ \\
 \vspace{2mm}
-y_+& x_+ & ict \\
\end{array}
\right]\cdot
\left[
  \begin{array}{ccc}
  \vspace{0.7mm}
    \dot{\breve{{\psi}}}_x^+\\
    \vspace{0.7mm}
    \dot{\breve{{\psi}}}_y^+ \\
    \vspace{0.7mm}
    \dot{\breve{{\psi}}}_z^+\\
  \end{array}
\right]=s_+\left[
  \begin{array}{ccc}
  \vspace{2mm}
    \eta_x^+ \\
    \vspace{2mm}
    \eta_y^+ \\
    \vspace{2mm}
    \eta_z^+\\
  \end{array}
\right],\quad
$$
and, respectively,
\begin{eqnarray}
\label{4.01a}
\left[
  \begin{array}{ccc}
\vspace{2mm}
    ict & z_- &- y_-\\
\vspace{2mm}
    -z_- & ict & x_- \\
\vspace{2mm}
    y_-& -x_- & ict \\
  \end{array}
\right]\cdot
\left[
\begin{array}{ccc}
\vspace{0.7mm}
\dot{\breve{{\psi}}}_x^-\\
\vspace{0.7mm}
\dot{\breve{{\psi}}}_y^- \\
\vspace{0.7mm}
\dot{\breve{{\psi}}}_z^-\\
  \end{array}
\right]=s_-\left[
  \begin{array}{ccc}
\vspace{2mm}
\eta_x^- \\
\vspace{2mm}
\eta_y^- \\
\vspace{2mm}
\eta_z^-\\
  \end{array}
\right],
\end{eqnarray}
where $ \dot{\breve{{\psi}}}_\sigma^\varsigma=\partial
\breve{{\psi}}_\sigma^\varsigma/\partial s_\varsigma$, in addition, the following notations
are made $\varsigma=(+,-)$ and $\sigma=(x,y,z).$

In addition, from (\ref{1.01}) we can find the following two relations between the
stochastic components of the wave function:
\begin{eqnarray}
  x_+ \dot{\breve{{\psi}}}^{+}_x+ y_+ \dot{\breve{{\psi}}}^{+}_y+ z_+\dot{\breve{{\psi}}}^{+}_z=0,
\nonumber\\
 x_-\dot{\breve{{\psi}}}^{-}_x+ y_- \dot{\breve{{\psi}}}^{-}_y+ z_-\dot{\breve{{\psi}}}^{-}_z=0.
\label{4.02}
\end{eqnarray}
For the further analytical study of the problem, it is important to reduce the system of equations
(\ref{4.01a}) to the canonical form:
\begin{eqnarray}
\dot{\breve{{\psi}}}^{+}_\sigma(s_+;\textbf{r}_+,t)=\bigl\{b^+_\sigma(\textbf{r}_+,t)+
d^+_\sigma(\textbf{r}_+,t)\bigr\}\bar{s}^{-1}_+\eta^+(s_+), \quad
\nonumber\\
\dot{\breve{{\psi}}}^{-}_\sigma(s_-;\textbf{r}_-,t)=\bigl\{b^-_\sigma(\textbf{r}_-,t)+
d^-_\sigma(\textbf{r}_-,t)\bigr\}\bar{s}^{-1}_-\eta^-(s_-), \quad
\label{4.02a}
\end{eqnarray}
where $\bar{s}_+=s_+/a_p$ and $\bar{s}_-=s_-/a_p$, in addition, the following notations are made:
 \begin{eqnarray}
b^\varsigma_x(\textbf{r}_\varsigma,t)=(z_\varsigma-y_\varsigma)/a_p,\qquad\qquad\qquad\qquad\,\,
\nonumber\\
b^\varsigma_y(\textbf{r}_\varsigma,t)=(x_\varsigma-z_\varsigma)/a_p,\qquad\qquad\qquad\qquad\,\,
\nonumber\\
b^\varsigma_z(\textbf{r}_\varsigma,t)=(y_\varsigma-x_\varsigma)/a_p,\qquad\qquad\qquad\qquad\,\,
\nonumber\\
d_x^\varsigma(\textbf{r}_\varsigma,t)=(c^2t^2-x^2_\varsigma-x_\varsigma y_\varsigma-x_\varsigma z_\varsigma)/(a_pct),
\nonumber\\
d^\varsigma_y(\textbf{r}_\varsigma,t)=(c^2t^2-y^2_\varsigma-x_\varsigma y_\varsigma-y_\varsigma z_\varsigma)/(a_pct),
\nonumber\\
d^\varsigma_z(\textbf{r}_\varsigma,t)=(c^2t^2-z^2_\varsigma-x_\varsigma z_\varsigma-y_\varsigma z_\varsigma)/(a_pct).
\label{4.02b}
\end{eqnarray}
Note that when deriving the equations (\ref{4.02b}) - (\ref{4.02a}) we assumed that the following
relations hold:
$$
\eta_x^\varsigma=\eta_y^\varsigma=\eta_z^\varsigma=\eta^\varsigma,\qquad  d\bar{s}_+=d\bar{s}_-=d\bar{s},
$$
which looks quite natural in the context of considered problem.

Now we will assume that the random generators satisfy the following correlation properties:
\begin{eqnarray}
\langle\eta(\bar{s})\rangle=0,\qquad \langle\eta(\bar{s})\eta(\bar{s}')\rangle=2\varepsilon\delta(\bar{s}-\bar{s}'),
\label{4.02ab}
\end{eqnarray}
where  $\eta(\bar{s})=\bar{s}^{-1}\eta^-(\bar{s})=\bar{s}^{-1}\eta^+(\bar{s})$ and
$\varepsilon= (\varepsilon^r+i\varepsilon^i)$ is a complex constant denoting the fluctuation
 power; in addition, we assume that
$\varepsilon_0=\varepsilon^r=\varepsilon^i.$

We can represent the joint probability distribution of fields in the form (see \cite{AshG}):
\begin{eqnarray}
\mathcal{P}(\{\breve{\psi}\},\bar{s};\textbf{r}_+,\textbf{r}_-,t)=\prod_{\varsigma,
\sigma }\bigl\langle\delta\bigl(\breve{\psi}_\sigma^\varsigma(\bar{s};\textbf{r}_\varsigma,t)-
\psi_{\sigma}^\varsigma\bigr)\bigr\rangle,
\label{4.02abc}
\end{eqnarray}
where  $\{{\breve{\psi}}\}=({\breve{\psi}}^+_x,...,{\breve{\psi}}^-_z)$ denotes a set of
fluctuations of vacuum fields and $\psi_{\sigma}^\varsigma=\breve{\psi}_{\sigma}^\varsigma(s)|_{t=0}$.
In the (\ref{4.02abc})  $\delta(\breve{\psi}_\sigma^\varsigma(s;\textbf{r}_\varsigma,t)-\psi_{\sigma}^\varsigma)$,
denotes the Dirac delta function generalized on a 6D-Hilbert space, in addition, by default we
will assume that the wave function is dimensionless, that is, it is multiplied by a constant value
$a_p^{3/2}$  (see (\ref{3.03a})).

Using the system of stochastic equations (\ref{4.02a}), for the conditional probability
the following second order partial differential equation can be obtained  (see \cite{Ashg1}):
\begin{eqnarray}
\Bigl\{\frac{\partial }{\partial\bar{s}}-\frac{1}{2} \sum_{\varsigma,\sigma }
\varepsilon_\sigma^\varsigma(\textbf{r}_\varsigma,t) \frac{\partial^2 }{\partial{\breve{\psi}}_\sigma^\varsigma\partial\bar{\breve{\psi}}_\sigma^\varsigma}\Bigr\}\mathcal{P}=0,
\label{4.03}
\end{eqnarray}
where $\bar{\breve{\psi}}_\sigma^\varsigma$ denotes the complex conjugate of the function
$\breve{\psi}_\sigma^\varsigma$ and $\varepsilon_\sigma^\varsigma(\textbf{r}_\varsigma,t)=
\varepsilon_0 [b_\sigma^\varsigma(\textbf{r}_\varsigma,t)+
d_\sigma^\varsigma(\textbf{r}_\varsigma,t)]^{2}$, which is  a dimensionless quantity and denotes fluctuations power.

Recall that in the equation (\ref{4.03}) the following notation is used:
$$
\frac{\partial^2 }{\partial{\breve{\psi}}_\sigma^\varsigma\partial\bar{\breve{\psi}}_\sigma^\varsigma}
=\frac{\partial^2 }{\partial\bigl[{\breve{\psi}}_\sigma^{\varsigma(r)}\bigr]^2}
+\frac{\partial^2 }{\partial\bigl[{\breve{\psi}}_\sigma^{\varsigma(i)}\bigr]^2},
$$
where  $\breve{\psi}_\sigma^{\varsigma(r)}=\emph{Re}(\breve{\psi}_\sigma^\varsigma)$ and
$\breve{\psi}_\sigma^{\varsigma(i)}=\emph{Im}(\breve{\psi}_\sigma^\varsigma) $.

For further it is convenient to represent the general solution of the equation
 (\ref{4.03}) in the following integral form:
\begin{widetext}
\begin{eqnarray}
\mathcal{P}(\{\breve{\psi}\},\bar{s}\,;\textbf{r}_+,\textbf{r}_-,t)=\int_{\Xi^6}
\mathcal{{P}}_0(\{\psi_0\})\prod_{\varsigma,\sigma}
\exp\biggl\{-\frac{(\psi_{\sigma}^\varsigma-{\breve{\psi}}_\sigma^\varsigma)(\bar{\psi}_{\sigma}^\varsigma
-\bar{{\breve{\psi}}}_\sigma^\varsigma) }{2\bar{s}\varepsilon_\sigma^\varsigma}\biggr\}
\frac{d \psi_{\sigma}^\varsigma}{ \sqrt{2\pi \bar{s}\varepsilon_\sigma^\varsigma}},\qquad
d\psi_{\sigma}^\varsigma=d\psi_{\sigma}^{\varsigma(r)}
d\psi_{\sigma}^{\varsigma(i)},
\label{4.03a}
\end{eqnarray}
\end{widetext}
where $\mathcal{{P}}_0(\{\psi_0\})$ denotes an initial condition of the equation (\ref{4.03})
at $t = 0$, before including of interaction with the random environment. Recall that the integration
over the 6D Hilbert space $\Xi^6$, ie by the spectrum, in accordance with the ergodic hypothesis, is
equivalent in this case to integration over the full 12D configuration space. Note that the
integration in (\ref{4.03a}) should be understood as an integral operator that acts on
each element of the matrix, which, as we shall show below, is the $\mathcal{{P}}_0(\{\psi_0\})$.

On the basis of physical considerations, as  an initial condition, we can choose the
probability density of the "\emph{ground state}" of a scalar field's particle, which
can be formed by two vector bosons with projections  of spin  $+1 $ and $-1 $, respectively.
It is obvious that the probability density $\mathcal{P}$ should be normalized on
the unit matrix of the size $I_3=3\times3$:
\begin{equation}
\int_{\Xi^6}\mathcal{P}(\{\bm\breve{\psi}\},\bar{s}\,;\textbf{r}_+,\textbf{r}_-,t)
\prod_{{\varsigma},{\sigma}}d{\breve{\psi}}_{{\sigma}}^{\varsigma}=I_3.
\label{4.03bt}
\end{equation}
Substituting $(\ref{4.03a})$ into $(\ref{4.03bt})$ and integrating over variables within
$\bigr[\breve{\psi}_{\sigma}^{\varsigma (r)},\, \breve{\psi}_{\sigma}^{\varsigma(i)}\bigl]
\in(-\infty,+\infty)$, we get the following condition for
the density matrix $\mathcal{{P}}_0(\{\psi_0\})$:
\begin{equation}
\int_{\Xi^6}\mathcal{{P}}_0(\{\psi_0\})
\prod_{\varsigma,\sigma}d{\psi}_{\sigma}^\varsigma=I_3.
\label{4.03btz}
\end{equation}

The wave function of a scalar field particle (\emph{scalar boson with spin-0}) can be represented
as the maximally  entangled state of two "\emph{ground states}" of the  vector field particles (EPR state \cite{Einst}):
\begin{equation}
\bm\psi^\updownarrow(\textbf{r}_+,\textbf{r}_-,0)= \frac{1}{\sqrt{2}}\bigl\{|\uparrow\rangle_1\otimes|
\downarrow\rangle_2-|\downarrow\rangle_1\otimes|\uparrow\rangle_2\bigr\},
\label{4.03b}
\end{equation}
where we make the following notation:
$|\uparrow\rangle_1=\bm\phi^+(\textbf{r}_+)$ and $|\downarrow\rangle_2=\bm\phi^-(\textbf{r}_-)$,
while  $|\downarrow \rangle_1=\bm\bar{\phi}^+(\textbf{r}_+)$ and $| \uparrow\rangle_2=
 \bm\bar{\phi}^-(\textbf{r}_-)$.
Note that the wave function (\ref{4.03b}) denotes one of Bell's entangled states of four possible \cite{Bell}.
Recall that in the equation (\ref{4.03b}) the symbol $ \otimes $ denotes the operation of a direct (tensor) product:
$$
\textbf{A}=|\uparrow\rangle_1\otimes| \downarrow\rangle_2 =
\left[
  \begin{array}{ccc}
  \vspace{1mm}
{\phi}_x^+\\
\vspace{1mm}
{\phi}_y^+\\
  \vspace{1mm}
{\phi}_z^+\\
  \end{array}
\right]
\left[
\begin{array}{ccc}
 {\phi}_x^-\,\,
 {\phi}_y^- \,\,
 {\phi}_z^-
  \end{array}
\right],\quad\,
$$
\begin{eqnarray}
\bar{\textbf{A}}=|\downarrow \rangle_1\otimes|\uparrow  \rangle_2 =
\left[
  \begin{array}{ccc}
  \vspace{1mm}
{\bar{\phi}}_x^+\\
\vspace{1mm}
{\bar{\phi}}_y^+\\
  \vspace{1mm}
{\bar{\phi}}_z^+\\
  \end{array}
\right]
\left[
\begin{array}{ccc}
 {\bar{\phi}}_x^-\,\,
 {\bar{\phi}}_y^- \,\,
 {\bar{\phi}}_z^-
  \end{array}
  \right],
  \label{A.01b}
\end{eqnarray}
where $\textbf{A}$ denotes the third-rank matrix, while $\bar{\textbf{A}}$  its complex conjugation.
The wave function of the boson of a scalar field can be represented in the form of a difference
of two matrices (see Appendix B):
\begin{eqnarray}
\bm\psi^\updownarrow(\textbf{r}_+,\textbf{r}_-, 0)=\frac{1}{\sqrt{2}}
\bigl\{\textbf{A}-\bar{\textbf{A}}\}=\frac{1}{\sqrt{2}}\,\textbf{B}=\quad
\nonumber\\
 \frac{1}{\sqrt{2}}\left[
  \begin{array}{ccc}
  \vspace{1mm}
B_{11}\,\,\,B_{12}\,\,\,B_{13}\\
\vspace{1mm}
B_{21}\,\,\,B_{22}\,\,\,B_{23}\\
 \vspace{1mm}
B_{31} \,\,\,B_{32}\,\,\,B_{33}\\
  \end{array}
\right]=\frac{1}{\sqrt{2}}
\left[
 \begin{array}{ccc}
  \vspace{1mm}
B_{11}\,\,\, 0\,\,\,\,0\,\,\\
\vspace{1mm}
0\quad B_{22}\,\,\,\, 0\,\\
 \vspace{1mm}
 \,\,0\quad\, 0\,\,\,\,\, B_{33}\\
  \end{array}
\right],\,\,
\label{4.03bz}
\end{eqnarray}
where the elements $B_{ij}=A_{ij}-\bar{A}_{ij}$ of the matrix $\textbf{B}$ are calculated explicitly:
\begin{eqnarray}
B_{11}=2i\bigl[\phi^{+(i)}_x\phi^{-(r)}_x-\phi^{+(r)}_x\phi^{-(i)}_x\bigr],
\nonumber\\
B_{22}=2i\bigl[\phi^{+(i)}_y\phi^{-(r)}_y-\phi^{+(r)}_y\phi^{-(i)}_y\bigr],
\nonumber\\
 B_{33}=2i\bigl[\phi^{+(i)}_z\phi^{-(r)}_z-\phi^{+(r)}_z\phi^{-(i)}_z\bigr].
\label{4.04}
\end{eqnarray}
Recall that the matrix elements $B_{ij}$ for $i\neq j$ are equal to zero, since the
localization domains of the functions occurring in them have no intersection.

The distribution of the probability in a scalar field's boson  before
switching-on  of the random environment has the form:
 \begin{eqnarray}
 \mathcal{{P}}_0(\{\psi_0\})=\frac{1}{C_0} \bigl|\bigr|\bm\psi^\updownarrow(\textbf{r}_+,\textbf{r}_-,0)
 \bigr|\bigr|^2 = \frac{1}{2C_0}\, \textbf{C},
\label{4.03bz}
\end{eqnarray}
where $C_0$ is the constant that will be found from the normalization condition
for the probability density on the matrix $ I_3 $, in addition:
$$\textbf{C}=\textbf{B}\cdot\textbf{B}=
\left[
 \begin{array}{ccc}
  \vspace{1mm}
C_{11}\,\,\,\,0\,\,\,\, 0 \\
\vspace{1mm}
0\quad C_{22}\,\,\,0\\
 \vspace{1mm}
 \,\,0\quad\, 0\,\,\,\,C_{33}\\
  \end{array}
\right],
$$
is a diagonal third rank matrix, the elements of which have the following form:
\begin{eqnarray}
C_{11}=B_{11}\bar{B}_{11},\quad C_{22}=B_{22}\bar{B}_{22},\quad
C_{33}=B_{33}\bar{B}_{33} ,
\nonumber\\
C_{13}=C_{12}=C_{23}=C_{32}=0.\qquad\qquad
\label{4.04}
\end{eqnarray}
Taking into account the fact that the spins of the vector bosons $\bm\phi^+$ and $\bm\phi^-$
lie on one axis but directed an opposite, for elements of the matrix $\textbf{C}$ we obtain
the following expressions:
 \begin{eqnarray}
C_{11}=4\bigl[\phi^{+(r)}_x\phi^{-(i)}_x\bigr]^2+4\bigl[\phi^{+(i)}_x\phi^{-(r)}_x\bigr]^2,
\nonumber\\
C_{22}=4\bigl[\phi^{+(r)}_y\phi^{-(i)}_y\bigr]^2+4\bigl[\phi^{+(i)}_y\phi^{-(r)}_y\bigr]^2,
\nonumber\\
C_{33}=4\bigl[\phi^{+(r)}_z\phi^{-(i)}_z\bigr]^2+4\bigl[\phi^{+(i)}_z\phi^{-(r)}_z\bigr]^2.
\label{4.04bzt}
\end{eqnarray}
\vspace{-1mm}
Substituting $(\ref{4.03bz})$ into $(\ref{4.03bt})$ and integrating, we can obtain:
\begin{eqnarray}
 \int_{\Xi^6}\mathcal{P}(\{\bm\psi\},\bar{s}\,;\textbf{r}_+,\textbf{r}_-,t)
\prod_{{\varsigma},{\sigma}}d{\breve{\psi}}_{{\sigma}}^{\varsigma}= \frac{1}{2C_0}\left[
 \begin{array}{ccc}
  \vspace{1mm}
 \overline{C}_{11}\,\,\,\,0\quad 0 \\
\vspace{1mm}
0\quad \overline{C}_{22}\,\,\,  0 \\
 \vspace{1mm}
 \,\,0\quad\, 0\,\,\,\,  \overline{C}_{33}\\
  \end{array}
\right],\nonumber\\
\label{4.05w}
\end{eqnarray}
where
$$
\overline{C}_{\sigma\sigma}=4\Bigl\{\overline{[\phi^{+(r)}_\sigma]^2}\,\cdot\overline{[\phi^{-(i)}_\sigma]^2}
+\overline{[\phi^{+(i)}_\sigma]^2}\,\cdot \overline{[\phi^{-(r)}_\sigma]^2}\,\Bigr\}.
$$
Recall that the terms $\overline{[\phi^{\varsigma(r)}_\sigma]^2}$ and
$\overline{[\phi^{\varsigma(i)}_\sigma]^2}$ denote the integrations of the corresponding matrix elements:
\begin{equation}
 \overline{[\phi^{+(i,r)}_\sigma]^2}=\int_{0}^1[\phi^{+(i,r)}_\sigma]^2d\phi^{+(i,r)}_\sigma=\frac{1}{3}.
\label{4.05c}
\end{equation}
Note that the fields $ \phi^{+(i, r)}_\sigma $ at integration are  used without constants
denoting the dimension and, therefore the range of their variation is  $[0,1]$
(see (\ref{3.03a})). Taking into account (\ref{4.05c}), it is easy to prove the equality
(\ref{4.03btz}). The normalization constant  in
this case is equal $C_0=\sqrt{3}(4/3)^4 $.

Finally, we can calculate the distribution of the probability density of fields in a
boson or the structure of a particle of a scalar field taking into account the influence
of a random environment.

Substituting (\ref{4.03bz}) into integral (\ref{4.03a}), and integrating the indefinite
integral we obtain:
\begin{eqnarray}
\mathcal{P}(\{\breve{\psi}\},\bar{s}\,;\textbf{r}_+,\textbf{r}_-,t)=\frac{1}{2C_0}
\left[
 \begin{array}{ccc}
  \vspace{1mm}
\widehat{C}_{11}\,\,\,\,0\,\,\,\,0 \\
\vspace{1mm}
0\quad \widehat{C}_{22}\,\,\,0\\
 \vspace{1mm}
 \,\,0\quad\,0\,\,\,\,\widehat{C}_{33}\\
  \end{array}
\right],
\label{4.06}
\end{eqnarray}
where
$$
\widehat{C}_{\sigma\sigma}=4\Bigl\{\widehat{\,\,[\phi^{+(r)}_\sigma]^2}\cdot\widehat{\,\,[\phi^{-(i)}_\sigma]^2}
+\widehat{\,\,[\phi^{+(i)}_\sigma]^2}\cdot\widehat{\,\,[\phi^{-(r)}_\sigma]^2}\Bigr\},
$$
and, in addition:
\begin{widetext}
\begin{equation}
\widehat{\,\,[\phi^{\varsigma(i,r)}_\sigma]^2}= -\frac{(\phi^{\varsigma(i,r)}_\sigma
+\breve{\psi}_{\sigma}^{\varsigma(i,r)})}{\sqrt{2\bar{s}\varepsilon_\sigma^\varsigma}}
\exp\biggl\{-\frac{\bigl(\phi^{\varsigma(i,r)}_\sigma-\breve{\psi}_{\sigma}^{\varsigma(i,r)}\bigr)^2}
{2\bar{s}\varepsilon_\sigma^\varsigma}\biggr\}+\frac{\sqrt{\pi}}{2}
\frac{[\breve{\psi}_{\sigma}^{\varsigma(i,r)}]^2}{2\bar{s}\varepsilon_\sigma^\varsigma}
\biggl\{1+erf\Bigl[\frac{\phi^{\varsigma(i,r)}_\sigma-\breve{\psi}_{\sigma}^{\varsigma(i,r)}}
{\sqrt{2\bar{s}\varepsilon_\sigma^\varsigma}}\Bigr]\biggr\}.
\label{4.06a}
\end{equation}
\end{widetext}
Recall that (\ref{4.06a}) is obtained up to a constant. As follows from the formula
(\ref{4.06a}), for the large fluctuations $||\breve{\psi}_{\sigma}^{\varsigma(i,r)})||\gg 1$, the value $\widehat{\,\,[\phi^{\varsigma(i,r)}_\sigma]^2}$ tends to zero, whereas  the function
$\phi^{\varsigma(i,r)}_\sigma$ is limited. It is obvious that with the help of expression
(\ref{4.06a}) we can construct all matrix elements of the diagonal matrix (\ref{4.06}).

Thus, as we have seen, the state of the Bose particle of the scalar field is quasistable,
in connection with which it can perform spontaneous transitions to other states both
inside and outside the vacuum.

\section{Conclusion}
Although no single fundamental scalar field has been experimentally observed so far,
 such fields play a key role in the constructions of modern theoretical physics.
There are important hypothetical scalar fields, for example the Higgs field for the \emph{Standard
Model} (SM), presence of each is also necessary for the completeness of the classification of
fundamental fields theory, including new physical theories, such as, for example, the \emph{String Theory} (SM).
Despite the great progress in the representations of modern particle theory within the framework
of the SM, it  does not give a clear explanation of a number of fundamental questions of the modern physics, such as
"\emph{What is dark matter}?" or  "\emph{What happened to the antimatter after the big bang}?"
and so on. Note that as modern astrophysical observations show, not less than 74 percent of the energy of
the universe is associated with a substance called \emph{dark energy}, which has no mass and whose
properties are not sufficiently studied and understood. Obviously, this substance must be connected
to \emph{quantum vacuum} or just be a QV itself. Note that in present-day understanding of what
is called the vacuum state or the quantum vacuum, it is "by no means a simple empty space".
In the vacuum state, electromagnetic waves and particles continuously appear and disappear,
so that on the average their number is zero.  It is obvious that these fluctuating or flickering
fields arise as a result of spontaneous decays of quasi-stable states of some massless and structured
field, very inert to any external interactions.

The main goal of this work was the theoretical substantiation of the possibility of
forming an uncharged  massless particle with zero spin in the region of negative energies.
As the basic equation describing in the coordinate representation a massless
particle with spin 1, we considered an equation of the Weyl type for neutrinos
(\ref{1.01a}). Assuming that on the scale of a particle of a vector field 4D
space-time is pseudo-Euclidean, ie of the Minkowski type (\ref{1.02k}), we determined the conditions
(\ref{1.02zt}) under which the existence of a structural particle in the form of 2D
string (\emph{brane}) is possible in the region of negative energies (see (\ref{3.03a})).
It is proved that the \emph{brane} is a stable only in the \emph{ground state}. The Bose
particle is described by a wave vector consisting of three components, each of which
characterizes the electrical and magnetic properties of the particle of a given
projection and is localized in the corresponding plane (see Fig. 1).

 A very important question, namely, what is the value of the parameter $ a_p$ (see (\ref{3.05at})),
 which characterizes the spatial size of the \emph{brane}, remains open within the framework
of the developed approach. Apparently, we can get a clear answer to this question with the
help of a series of experiments that we plan carry out in the near future. In particular, if the
value of the constant $a_p$ will be significantly different, from the Planck length $l_P$, then it will be
necessary to introduce a new fundamental constant characterizing the \emph{brane} size.

In the ensemble of
\emph{branes}, random interactions occur that lead to the formation of maximally entangled
massless particles (entangled-brane) with spin-0. We assume that these states are the most
stable and, accordingly, their weight in the balance of fields will be the greatest.
The process of formation of \emph{e-branes} and their Bose condensation is studied in detail
 in the framework of stochastic equations (\ref{4.02a}). The equation is obtained for the density of the
\emph{e-brane} wave state (see (\ref{4.06a})), by means of which it is possible to
calculate spontaneous transitions to various free and bound states, including
states with the formation of massive particle  and antiparticle, for example,
an electron-positron pair.

Thus, the conducted studies allow us to speak about the structure and properties of
QV and, accordingly, about the structure and properties of the "empty" space-time.
In the end, we note that as preliminary studies show that the properties of
space-time measurably can be changed at relatively low external fields, which
will be extremely important for future technologies.

\section{Appendix}

\subsection{}
Using the systems of equations (\ref{1.02k})-(\ref{1.02ta})  and carrying out a
similar calculations for vacuum fields consisting of particles with projections of spin -1, we can
obtain the system of following equations:
\begin{eqnarray}
 \bigl\{\square+ [i(c_{,y}-c_{,z})+c_{,\tau}c^{-1}]c^{-2}\partial_\tau\bigr\}\psi^{-}_x=0,
\nonumber\\
\bigl\{\square+[i(c_{,z}-c_{,x})+c_{,\tau}c^{-1}]c^{-2}\partial_\tau\bigr\}\psi^{-}_y=0,
\nonumber\\
\bigl\{\square+[i(c_{,x}-c_{,y})+c_{,\tau}c^{-1}]c^{-2}\partial_\tau\bigr\}\psi^{-}_z=0.
 \label{A.l}
\end{eqnarray}
In (\ref{A.l}) substituting the solutions of the equations in the form:
 \begin{equation}
 \psi_\sigma^-(\textbf{r},\tau)=\exp{\Bigl( \frac{\mathcal{E}_\sigma \tau}{\hbar}\Bigr)}\phi_\sigma^-(\textbf{r}),\qquad
 \sigma=x,y,z,
\label{A.2}
\end{equation}
we obtain the following stationary equations:
\begin{eqnarray}
\Bigl\{\triangle + \lambda\Bigl[-\lambda-\frac{1}{r}+i\frac{y-z}{ r^2}\Bigr]\Bigr\}\phi_x^-(\textbf{r})
=0,
\nonumber\\
\Bigl\{\triangle + \lambda\Bigl[-\lambda-\frac{1}{r}+i\frac{z-x}{ r^2}\Bigr]\Bigr\}\phi_y^-(\textbf{r})
=0,
\nonumber\\
\Bigl\{\triangle +\lambda\Bigl[-\lambda-\frac{1}{r}-i\frac{x-y}{r^2}\Bigr]\Bigr\}
 \phi_z^+(\textbf{r})=0,
\label{A.2a}
\end{eqnarray}
 where $\lambda=(\mathcal{E}/c\hbar)<0$.

The solution of the system of equations is carried out in a similar way, as for fields $\phi^+(\textbf{r})$
(see Eq.s (\ref{3.020})-(\ref{3.02b})). In particular, calculations show that the components of
a vector boson with a spin projection -1 are localized on the planes, $\bigl[\phi_x^{-(r)}(Y,-Z),\,\,
\phi_x^{-(i)}(-Y,Z)\bigr],\quad\bigl[\phi_y^{-(r)}(X,-Z),$ $\,\,\phi_x^{-(i)}(-X,Z)\bigr]$ and
$\bigl[\phi_z^{-(r)}(X,-Y),\,\,\phi_x^{-(i)}(-X,Y)\bigr]$, respectively.

\subsection{}

The direct product between two vector states is represented in the form:
$$
|\uparrow\rangle_1\otimes| \downarrow\rangle_2 =
\left[
  \begin{array}{ccc}
  \vspace{1mm}
{\psi}_x^+\\
\vspace{1mm}
{\psi}_y^+\\
  \vspace{1mm}
{\psi}_z^+\\
  \end{array}
\right]
\left[
\begin{array}{ccc}
 {\psi}_x^-\,\,
 {\psi}_y^- \,\,
 {\psi}_z^-
  \end{array}
\right] =
$$
$$
\left[
  \begin{array}{ccc}
  \vspace{1mm}
{\psi}_x^+{\psi}_x^- \,\,\, {\psi}_x^+{\psi}_y^-\,\,\,{\psi}_x^+{\psi}_z^-\\
\vspace{1mm}
 {\psi}_y^+{\psi}_x^- \,\,\, {\psi}_y^+{\psi}_y^-\,\,\,{\psi}_y^+{\psi}_z^- \\
 \vspace{1mm}
  {\psi}_z^+{\psi}_x^- \,\,\, {\psi}_z^+{\psi}_y^-\,\,\,{\psi}_z^+{\psi}_z^-\\
  \end{array}
\right]=
\left[
  \begin{array}{ccc}
  \vspace{1mm}
A_{11}\,\,\,A_{12}\,\,\,A_{13}\\
\vspace{1mm}
 A_{21}\,\,\,A_{22}\,\,\,A_{23}\\
 \vspace{1mm}
  A_{31} \,\,\,A_{32}\,\,\,A_{33}\\
  \end{array}
\right],
$$
and, respectively, for their complex conjugation:
\begin{eqnarray}
\bar{\textbf{A}}=|\downarrow \rangle_1\otimes|\uparrow  \rangle_2 =
\left[
  \begin{array}{ccc}
  \vspace{1mm}
{\bar{\psi}}_x^+\\
\vspace{1mm}
{\bar{\psi}}_y^+\\
  \vspace{1mm}
{\bar{\psi}}_z^+\\
  \end{array}
\right]
\left[
\begin{array}{ccc}
 {\bar{\psi}}_x^-\,\,
 {\bar{\psi}}_y^- \,\,
 {\bar{\psi}}_z^-
  \end{array}
  \right]= \quad
  \nonumber\\
 \left[
  \begin{array}{ccc}
  \vspace{1mm}
{\bar{\psi}}_x^+{\bar{\psi}}_x^- \,\,\, {\bar{\psi}}_x^+{\bar{\psi}}_y^-\,\,\,{\bar{\psi}}_x^+{\bar{\psi}}_z^-\\
\vspace{1mm}
 {\bar{\psi}}_y^+{\bar{\psi}}_x^- \,\,\, {\bar{\psi}}_y^+{\bar{\psi}}_y^-\,\,\,{\bar{\psi}}_y^+{\bar{\psi}}_z^-\\
 \vspace{1mm}
  {\bar{\psi}}_z^+{\bar{\psi}}_x^- \,\,\, {\bar{\psi}}_z^+{\bar{\psi}}_y^-\,\,\,{\bar{\psi}}_z^+{\bar{\psi}}_z^-\\
  \end{array}
\right]=
\left[
  \begin{array}{ccc}
  \vspace{1mm}
\bar{A}_{11}\,\,\,\bar{A}_{12}\,\,\,\bar{A}_{13}\\
\vspace{1mm}
 \bar{A}_{21}\,\,\,\bar{A}_{22}\,\,\,\bar{A}_{23}\\
 \vspace{1mm}
  \bar{A}_{31} \,\,\,\bar{A}_{32}\,\,\,\bar{A}_{33}\\
  \end{array}
\right].
\label{A.01b}
\vspace{7mm}
\end{eqnarray}

\end{document}